\documentclass[pre]{revtex4}
\usepackage{graphicx}
\begin{document}
\title{Generalized Einstein Relation for Brownian Motion in Tilted Periodic Potential}
\author{Hidetsugu Sakaguchi}
\address{Department of Applied Science for Electronics and Materials,\\
Interdisciplinary Graduate School of Engineering Sciences,\\
Kyushu University, Kasuga, Fukuoka 816-8580, Japan
}
\begin{abstract}
A generalized Einstein relation is studied for Brownian motion in a tilted potential. The exact form of the diffusion constant of the Brownian motion is compared with the generalized Einstein relation. The generalized Einstein relation is a good approximation in a parameter range where the Brownian motion exhibits stepwise motion.   
\end{abstract}
\maketitle
\section{Introduction}
Recently, the fluctuation-dissipation theorem in nonequilibrium systems 
has again been studied by several authors.~\cite{rf:1,rf:2} The Einstein relation is the simplest form of the fluctuation-dissipation theorem.
In a recent paper, we have studied a rotary Brownian motion in a simple tight-coupling model of molecular motors, and found a generalized Einstein relation between the diffusion coefficient $D$ and the response coefficient $\gamma$ for the  static external force $F$.~\cite{rf:3} The relation is derived on the basis of the fluctuation theorem~\cite{rf:4} and  an approximation of the stepwise Brownian motion by a random walk. In this paper, we check the generalized Einstein relation for Brownian motion in a tilted periodic potential. Thermal diffusion in a tilted periodic potential has a prominent role in various systems such as Josephson junctions~\cite{rf:5}, systems for diffusion on crystal surfaces~\cite{rf:6} and noisy limit cycle oscillators~\cite{rf:7}. It is a simple and typical problem in nonlinear-nonequilibrium physics. 

\section{Brownian Motion in  Tilted Periodic Potential and  Generalized Einstein Relation}
We study an overdamped one-dimensional Brownian motion in a tilted sinusoidal potential. The model equation is written as 
\begin{equation}
\frac{dx}{dt}=-B\sin x+F+\xi(t),
\end{equation}
where $F$ is the static tilting force, and $\xi$ is the Gaussian white noise satisfying $\langle \xi(t)\cdot \xi(t^{\prime})\rangle=2T\delta(t-t^{\prime})$.
The average velocity $v=\langle \dot{x}\rangle$ and the effective diffusion coefficient $D$ are explicitly calculated as \cite{rf:8}
\begin{equation}
v=\frac{1-\exp(-2\pi F/T)}{\int_0^{2\pi}dxI_{+}(x)/(2\pi)},
\end{equation}
and
\begin{equation}
D=T\frac{\int_0^{2\pi}dxI_{+}(x)^2I_{-}(x)/(2\pi)}{[\int_0^{2\pi}dxI_{+}(x)/(2\pi)]^3},
\end{equation}
where 
\begin{equation}
I_{\pm}(x)=\frac{1}{T}\int_0^{2\pi}dy\exp[\{\mp B\cos x\pm B\cos(x\mp y)-yF\}/T].
\end{equation}

\begin{figure}[htb]
\begin{center}
\includegraphics[width=10cm]{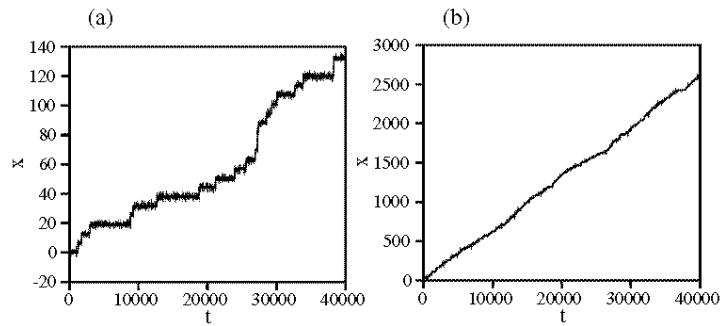}
\end{center}
\caption{Time evolution of $x(t)$  (a) for $F=0.3$ and (b) $F=0.6$ at $B=1$ and $T=0.2$ by eq.~(1).}
\label{fig:1}
\end{figure}
We show a result of a Langevin simulation of eq.~(1). Figure 1 displays time evolutions for (a) $F=0.3$ and (b) $F=0.6$ at $B=1$ and $T=0.2$. There are potential minima at $x=x_0+2\pi n$, where $x_0=\sin^{-1}(F/B)<\pi/2$, in the tilted potential $U(x)=-Fx-B\cos x$ for $F<B$. The interval between the neighboring potential minima is $2\pi$.
The Brownian motion exhibits stepwise motion clearly seen in Fig.~1(a) for smaller values of $F$, because a Brownian particle is trapped near the potential minima.  As $F$ or $D$ increases, stepwise motion becomes unclear as Fig.~1(b). 
The stepwise Brownian motion might be approximated by a random walk, in which the particle 
jumps randomly between the potential minima at discrete time steps $m\Delta t$.  The transition probability during one step $\Delta t$ from the $n$th potential minimum  to the $(n\pm 1)$th minimum is denoted by $p_{\pm}$. The probability that the particle stays at the same site is denoted as $p_{0}=1-p_{+}-p_{-}$. 
The average velocity and diffusion coefficient of the random walk are calculated as \cite{rf:3} 
\begin{eqnarray}
v&=&2\pi(p_{+}+p_{-})/\Delta t,\nonumber\\
D&=&(2\pi)^2(p_{+}+p_{-})/(2\Delta t),
\end{eqnarray}
if $\Delta t$ is sufficiently small.  The diffusion coefficient is rewritten using $v$ as 
\begin{equation}
D=\frac{2\pi v(p_{+}-p_{-})}{2(p_{+}-p_{-})}.
\end{equation}
On the other hand, the fluctuation theorem tells that the ratio $p_{+}/p_{-}$ is equal to $\exp(2\pi F/T)$, because the entropy production for the transition  from the $n$th site to the $(n+1)$th site is $\sigma=2\pi F/T$.  
The substitution of this relation to eq.~(6) yields
\begin{equation}
D=\frac{2\pi v}{2{\rm tanh}(\pi F/T)}=\gamma(F) T_{eff},
\end{equation}
where $\gamma(F)=v/F$ and the effective temperature $T_{eff}=(\pi F)/{\rm tanh}(\pi F/T)$. This is the generalized Einstein relation. For sufficiently small $F$, $\gamma(F)=\lim_{F\rightarrow 0}dv/dF$ and $T_{eff}=T$; thus, the standard Einstein relation $D=\gamma(0) T$ is recovered. The Einstein relation is the simplest form of the fluctuation-dissipation theorem. 
The generalized Einstein relation is satisfied even in a nonlinear regime of a relation of $v$ and $F$. It does not depend on the detailed form of the potential. The effective temperature $T_{eff}$ is larger than $T$, because $z/{\rm tanh} z\ge 1$. 

\begin{figure}[htb]
\begin{center}
\includegraphics[width=14cm]{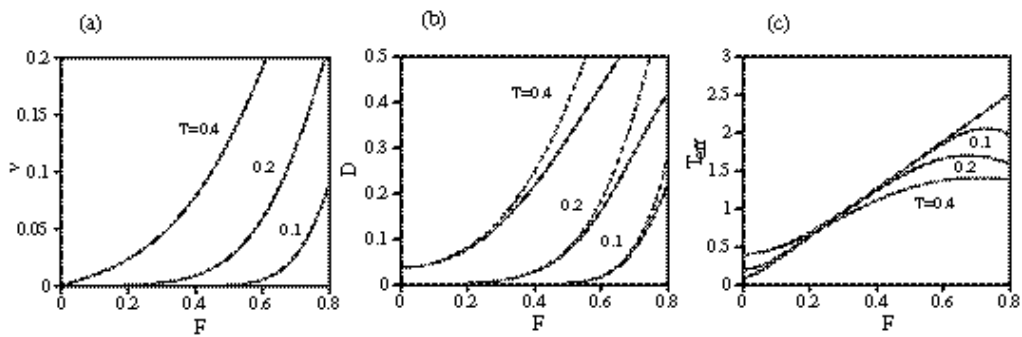}
\end{center}
\caption{(a) Relation of $v$ vs $F$ for $T=0.1,0.2$ and $T=0.4$ at $B=1$. (b) Relation of $D$ vs $F$. The solid curve denotes eq.~(3) and the dashed curve denotes eq.~(7). (c) Effective temperature $T_{eff}$ by $D/\gamma(F)$ (solid curve) and $T_{eff}=(\pi F)/\tanh(\pi F/T)$ (dashed curve).}
\label{fig:2}
\end{figure}
Figure 2(a) shows the average velocity $v$ as a function of $F$ for $T=0.1,0.2$ and 0.4 by eq.~(2) with $B=1$.
The nonlinear relation of $v$ and $F$ is clearly seen. A comparison of the exact  $D$ using eq.~(3) and (7) is shown in Fig.~2(b) for $T=0.1,0.2$ and $0.4$. 
 The generalized Einstein relation is a good approximation for relatively small $F$. Figure 2(c) shows the effective temperature obtained by the relation $T_{eff}=D/\gamma(F)$ and $T_{eff}=(\pi F)/\tanh(\pi F/T)$. Both the effective temperature starts from $T_{eff}(0)=T$ at $F=0$. The effective temperature by $T_{eff}=(\pi F)/\tanh(\pi F/T)$ increases monotonically, but the exact $T_{eff}=D/\gamma(F)$ has a peak and decreases for sufficiently large $F$. However, the approximation using $T_{eff}=(\pi F)/\tanh(\pi F/T)$ is very good for a relatively small $F$. The reason for the disagreement is that the stepwise motion becomes unclear and that the approximation by the random walk becomes worse for larger $F$ or larger $T$.

\section{Approximate Derivation of Generalized Einstein Relation}
The above derivation of the generalized Einstein relation is simple and general, but not very analytical. In this section, the Einstein relation is exactly derived, and the generalized Einstein relation is approximately derived directly from eqs.~(2) and (3).  For $F=0$, $I_{\pm}$ becomes
\begin{equation}
I_{\pm}(x)=\frac{1}{T}\exp\{\mp B\cos x/T\}I_{\pm 0},
\end{equation}
where $I_{\pm0}=\int_0^{2\pi}dy\exp\{\pm B\cos y/T\}$.
 For sufficiently small $F$, $v$ and $D$ are given by
\begin{eqnarray}
v&=&\frac{2\pi F/T}{I_{+0}I_{-0}/(2\pi T)},\nonumber\\
D&=&\frac{T I_{+0}^2I_{-0}^2/(2\pi T^3)}{\{I_{+0}I_{-0}/(2\pi T)\}^3}=\frac{4\pi^2 T}{I_{+0}I_{-0}}.
\end{eqnarray}
The ratio $D/v$ is therefore equal to $D/v=T/F$. This is the Einstein relation.

When $F$ ($F<B$) is not so small, and $B/T$ is sufficiently large, the Brownian particle is strongly trapped near the potential minimum $x=x_0+2\pi n$. 
The integral $I_{+}(x)$ in eq.~(4) is rewritten as
\[I_{+}(x)=\frac{1}{T}\exp\{(-B\cos x-xF)/T\}\int_{x-2\pi}^{x}dy\exp\{(B\cos y+yF)/T\}\]
with a change of variable $x-y\rightarrow y$. 
The integrand of $I_{+}$ is almost zero except near the potential minimum.  
If the integrand of $I_+$ is approximated using a Gaussian function near the potential minimum $y=x_0-2\pi$ (for $x<x_0$) or $y=x_0$ (for $x>x_0$), the integral $I_{+}$ is evaluated as  
\begin{eqnarray}
I_{+}(x)&\sim&  \sqrt{\frac{2\pi T}{B\cos x_0}}\frac{1}{T}\exp\{(-B\cos x-xF)/T\}\exp\{(B\cos x_0+(x_0-2\pi)F)/T\} \;\;\;\;\;\; {\rm for}\;\; x<x_0,\nonumber\\
&\sim&  \sqrt{\frac{2\pi T}{B\cos x_0}}\frac{1}{T}\exp\{(-B\cos x-xF)/T\}\exp\{(B\cos x_0+x_0F)/T\} \;\;\;\;\;\; {\rm for}\;\; x>x_0.\nonumber
\end{eqnarray}
Similarly, $I_{-}$ is evaluated as
\begin{eqnarray}
I_{-}(x)&=&\frac{1}{T}\exp\{(B\cos x+xF)/T\}\int_{x}^{x+2\pi}dy\exp\{(-B\cos y-yF)/T\}\nonumber\\
&\sim&  \sqrt{\frac{2\pi T}{B\cos x_0}}\frac{1}{T}\exp\{(B\cos x+xF)/T\}\exp\{(B\cos x_0+(x_0-\pi)F)/T\} \;\;\;\;\;\; {\rm for}\;\; x<\pi-x_0,\nonumber\\
&\sim&  \sqrt{\frac{2\pi T}{B\cos x_0}}\frac{1}{T}\exp\{(B\cos x+xF)/T\}\exp\{(B\cos x_0+(x_0-3\pi)F)/T\} \;\;\;\;\;\; {\rm for}\;\; x>\pi-x_0,\nonumber
\end{eqnarray}
because the integrand of $I_{-}$ takes a maximum at $y=\pi-x_0$ (for $x<\pi-x_0$) or $y=3\pi-x_0$ (for $x>\pi-x_0$).
The integrals in eqs.~(2) and (3) are calculated as
\begin{eqnarray}
\int_0^{2\pi}dx I_{+}(x)&=&\left (\frac{2\pi T}{B\cos x_0}\right )^{1/2}\frac{1}{2\pi T} [\int_0^{x_0}dx\exp\{(-B\cos x-xF)/T\}\exp\{(B\cos x_0+(x_0-2\pi)F)/T\}
\nonumber\\
&+&\int_{x_0}^{2\pi}dx\exp\{(-B\cos x-xF)/T\}\exp\{(B\cos x_0+x_0F)/T\} ]\nonumber\\ 
&\sim &\frac{1}{B\cos x_0}\exp\{(2B\cos x_0+(2x_0-\pi)F)/T\},\nonumber\\
\int_0^{2\pi}dx I_{+}^2I_{-}&=&\left (\frac{2\pi T}{B\cos x_0}\right )^{3/2}\frac{1}{2\pi T^3}[\int_0^{x_0}dx\exp\{(-B\cos x-xF)/T\}\exp\{(3B\cos x_0+(3x_0-3\pi)F)/T\}\nonumber\\
&+&\int_{x_0}^{\pi-x_0}dx\exp\{(-B\cos x-xF)/T\}\exp\{(3B\cos x_0+(3x_0-\pi)F)/T\}\nonumber\\
&+&\int_{\pi-x_0}^{2\pi}dx\exp\{(-B\cos x-xF)/T\}\exp\{(3B\cos x_0+(3x_0-3\pi)F)/T\} ] \nonumber\\ 
&\sim &\frac{1}{(B\cos x_0)^2}\frac{2\pi}{2T}\exp\{(4B\cos x_0+4x_0F)/T\}\{\exp(-2\pi F/T)+\exp(-4\pi F/T)\}.\nonumber
\end{eqnarray}
Here, the integral is evaluated at $x=\pi-x_0$. The ratio $D/v$ is given by 
\begin{equation}
\frac{D}{v}=\frac{T\int_0^{2\pi}dx I_{+}^2I_{-}/(2\pi)}{\{1-\exp(-2\pi F/T)\}\left(\int_0^{2\pi}dxI_+(x)/(2\pi)\right )^2}=\frac{2\pi}{2\tanh(\pi F/T)}.
\end{equation}
Thus, the generalized Einstein relation was approximately derived. 
\section{Summary}
We have confirmed that the generalized Einstein relation is satisfied for Brownian motion in a tilted potential, if the random walker exhibits stepwise motion. The diffusion coefficient by eq.~(3) is exact, but it depends on the detailed form of the potential. The generalized Einstein relation has a simple and general form, which is satisfied even in a nonlinear and nonequilibrium regime. 

\end{document}